# High-throughput imaging measurements of thermoelectric figure of merit

Abdulkareem Alasli[a,*], Asuka Miura[b,*,†], Ryo Iguchi[b], Hosei Nagano[a] and Ken-ichi Uchida[b,c,d]

[a]Department of Mechanical Systems Engineering, Nagoya University, Nagoya 464-8601, Japan;
[b]National Institute for Materials Science, Tsukuba 305-0047, Japan;
[c]Institute for Materials Research, Tohoku University, Sendai 980-8577, Japan;
[d]Center for Spintronics Research Network, Tohoku University, Sendai 980-8577, Japan

[*]These authors contributed equally to this work.
[†]Present address: Integrated Research for Energy and Environment Advanced Technology, Kyushu Institute of Technology, Kitakyushu, Fukuoka 804-8550, Japan

CONTACT Ken-ichi Uchida UCHIDA.Kenichi@nims.go.jp

We demonstrate a method for the simultaneous determination of the thermoelectric figure of merit of multiple martials by means of the lock-in thermography (LIT) technique. This method is based on the thermal analyses of the transient temperature distribution induced by the Peltier effect and Joule heating, which enables high-throughput estimation of the thermal diffusivity, thermal conductivity, volumetric heat capacity, Seebeck or Peltier coefficient of the materials. The LIT-based approach has high reproducibility and reliability because it offers sensitive noncontact temperature measurements and does not require the installation of an external heater. By performing the same measurements and analyses with applying an external magnetic field, the magnetic field and/or magnetization dependences of the Seebeck or Peltier coefficient and thermal conductivity can be determined simultaneously. We demonstrate the validity of this method by using several ferromagnetic metals (Ni, $Ni_{95}Pt_5$, and Fe) and a nonmagnetic metal (Ti). The proposed method will be useful for materials research in thermoelectrics and spin caloritronics and for investigation of magneto-thermal and magneto-thermoelectric transport properties.

## 1. Introduction

The Seebeck and Peltier effects refer to the conversion of heat into electricity and vice versa [1–3]. Since the discovery of these fundamental thermoelectric phenomena, many developments in physics, materials science, and applications have been achieved in order to fulfill the increasing global demands for energy [4–6]. In this context, accurate evaluation of the performance of thermoelectric materials is of great importance. The dimensionless figure of merit $ZT$ is a well-known parameter representing the performance of thermoelectric materials [7]. $ZT$ can be measured directly or derived from the transport properties of a material since it is defined as

$$ZT = \frac{S^2 \sigma}{\kappa} T \left( = \frac{\Pi^2 \sigma}{\kappa} \frac{1}{T} \right), \qquad (1)$$

where $S = \Pi/T$, $\sigma$, $\kappa$, and $T$ are the Seebeck coefficient with $\Pi$ being the Peltier coefficient, electrical conductivity, thermal conductivity, and absolute temperature, respectively [3]. The direct measurement of $ZT$ is realized by the so-called Harman method, which determines $ZT$ from the voltage responses of a material to DC and AC charge currents with a single apparatus [8,9]. The accuracy of the Harman method, however, can be affected by unavoidable electrical contact resistance and heat leakage through measurement probes [10]. Therefore, $ZT$ is often estimated by measuring $S$, $\sigma$, and $\kappa$ independently, although such measurements require a wide range of time-consuming experiments and lead to larger combined uncertainties [11].

In this article, we propose a different approach for determining $ZT$ based on the lock-in thermography (LIT). LIT is an active thermal imaging technique that allows the detection of the thermal response in a sample to a periodic external perturbation with high temperature resolution [12]. Although LIT was originally developed for nondestructive testing of electronic components [13], it has shown notable performance in the measurements of various properties including the Peltier coefficient [14], thermal diffusivity [15], magnetocaloric effects [16,17], magneto-thermoelectric effects [18–22], and thermo-spin effects [23,24]. Here, we show that LIT also enables simultaneous measurements of $S$ and $\kappa$, allowing the rapid derivation of $ZT$ with a single apparatus. Since LIT is a noncontact thermal imaging technique, the thermoelectric and thermal transport properties of multiple materials can be measured at the same time without the disturbance of heat leakage through temperature probes. The proposed method can be easily extended to the measurements under magnetic fields, enabling the investigation of magneto-thermal resistance [22] and magneto-thermoelectric effects.

## 2. Measurement and analysis procedures

The basic concept and principle of the LIT-based $ZT$ measurement method are illustrated in Figure 1. The setup consists of an infrared camera connected to a processing system and current source. A typical sample is a bar-shaped junction comprising thermoelectric and reference materials [Figure 1(a)]. The estimation of $S$ and $\kappa$ of the thermoelectric material is the target of this measurement, while the transport properties of the reference material are given parameters. When a charge current $\mathbf{J}_c$ is applied to the junction, the temperature change due to the Peltier effect (Joule heating) appears at the junction interface (in the bulk of the materials). LIT enables the imaging of the transient temperature modulation signals induced by the Peltier effect and Joule heating and the separation of them as follows. Under a square-wave-modulated AC charge current with the frequency $f$, amplitude $J_c$, and zero DC offset, an oscillating heat release and absorption due to the Peltier effect with the same $f$ and a constant heat release due to the Joule heating are induced [LIT measurement 1 in Figure 1(b)]. On the other hand, under a current with the frequency $f$, amplitude $J_c/2$, and DC offset $J_c/2$, oscillating Peltier and Joule-heating signals with the same $f$ are induced [LIT measurement 2 in Figure. 1(c)]. Consequently, in the LIT measurement 1, LIT outputs the lock-in amplitude $A_\Pi$ and phase $\phi_\Pi$ images of the Peltier-effect-induced temperature modulation through Fourier analysis [12], where the Joule-heating contribution is eliminated by extracting the first harmonic response of the temperature modulation. Here, the amplitude image shows the spatial distribution of the magnitude of the temperature modulation, while the phase image shows the time delay of the temperature modulation and heat flux direction. In contrast, in the LIT measurement 2, both the Peltier and Joule-heating contributions appear in LIT images. Thus, to extract the amplitude $A_J$ and phase $\phi_J$ images of the Joule-heating-induced temperature modulation, we subtract the half of the Peltier-effect-induced LIT signals obtained in the LIT measurement 1 from raw LIT signals in the LIT measurement 2, which cancels the Peltier contribution because of the absence of the Joule-heating contribution in the LIT measurement 1. Based on the thermal analyses of the obtained images, the parameters of interest can be obtained, as detailed below.

Figure 1(d) summarizes the modelling and sequence of the thermal analyses in our method. During the LIT measurements, the Peltier effect generates an oscillating heat current density $\mathbf{q} = \Pi \mathbf{j}_c$ in the thermoelectric and reference materials, where $\mathbf{j}_c$ is a current charge density. At the junction, $\mathbf{q}$ becomes discontinuous due to the difference in $\Pi$ between the thermoelectric and reference materials, which leads to finite heating or cooling $Q_\Pi$ ($= -\nabla \cdot \mathbf{q}$). The oscillating $Q_\Pi$ in time diffuses as heat waves from the junction into each material. The propagation of the heat waves is dominated by the thermal diffusivity $D = \kappa/(\rho c_p)$ with $\rho$ and $c_p$ respectively being the density and specific heat capacity. Due to the difference in the thermophysical properties, $Q_\Pi$ diffuses asymmetrically and the induced temperature modulation $\Delta T_\Pi$ decays exponentially along the bar [Figure 1(d)]. The $\Delta T_\Pi$ distribution in each material can be obtained by solving the one-dimensional heat equation along the $x$ direction for the infinitely long junction system, whose interface is located at $x = 0$ and the thermoelectric (reference) material spans $x < 0$ ($x > 0$) [25]. $\Delta T_\Pi$ is assumed to be continuous at the interface by neglecting the interfacial thermal resistance and to be zero at the system boundaries ($x \to \pm\infty$). Under the adiabatic condition, the amplitude $A_\Pi^{\text{TE(Ref)}}$ and phase delay $\phi_\Pi^{\text{TE(Ref)}}$ of the first harmonic of the steady periodic solution of $\Delta T_\Pi$ is given by

$$A_\Pi^{\text{TE(Ref)}}(x) = \sqrt{\frac{8}{\pi^3 f}} \frac{(\Pi_{\text{TE}} - \Pi_{\text{Ref}}) j_c}{(\kappa_{\text{Ref}}/\sqrt{D_{\text{Ref}}} + \kappa_{\text{TE}}/\sqrt{D_{\text{TE}}})} e^{+(-)\sqrt{\frac{D_{\text{TE(Ref)}}}{\pi f}} x}, \qquad (2)$$

$$\phi_\Pi^{\text{TE(Ref)}}(x) = -(+)\sqrt{\frac{D_{\text{TE(Ref)}}}{\pi f}} x + \frac{\pi}{4}, \qquad (3)$$

where $j_c$ denotes the square-wave amplitude of the charge current density and the superscripts or subscripts TE and Ref refer to the thermoelectric and reference materials, respectively. Similarly, the Joule-heating-induced temperature modulation $\Delta T_J$ can be obtained by solving the heat balance equation under the condition of negligible heat losses [26,27]; the amplitude $A_J^{\text{TE(Ref)}}$ and phase delay $\phi_J$ of $\Delta T_J$ are given by

$$A_J^{\text{TE(Ref)}} = \frac{D_{\text{TE(Ref)}} j_c^2}{\pi^2 f \sigma_{\text{TE(Ref)}} \kappa_{\text{TE(Ref)}}}, \qquad (4)$$

$$\phi_J = \frac{\pi}{2} \qquad (5)$$

Here, the heat release or absorption due to the temperature dependence of $S$, i.e., the Thomson effect, is neglected because its contribution in metals is usually small [28,29]. Accordingly, by analyzing the LIT images using Equations (2)-(4), $D$, $\kappa$, $\Pi$, and $ZT$ of the thermoelectric material can be determined [Figure 1(d)]. $D$ can be derived by fitting observed $A_\Pi$ profiles with Equation (2) (referred to as $D_A$) and $\phi_\Pi$ profiles with Equation (3) (referred to as $D_\phi$). $\kappa$ can be then determined from $A_J$ by using the $\sigma$ values measured separately. Subsequently, $\Pi$ and $S$ can be calculated from $A_\Pi$ with respect to the reference material. $ZT$ is then estimated based on Equation (1).

To demonstrate the usability of the proposed method, we measured the thermal responses from four different materials simultaneously by means of LIT. We used polycrystalline Ni, Ni$_{95}$Pt$_5$, Fe, and Ti slabs as thermoelectric materials and a Cu slab as a reference material. The Ni, Fe, Ti, and Cu slabs are commercially available from the Nilaco Corporation, Japan, and the Ni$_{95}$Pt$_5$ slab, prepared by a melting method with rapid cooling, is available from Kojundo Chemical Laboratory Co., Ltd., Japan. The Ni/Cu, Ni$_{95}$Pt$_5$/Cu, Fe/Cu, and Ti/Cu junctions were prepared by a spark plasma sintering method at 800 °C under a pressure of 30 MPa for 5 minutes in a vacuum. The junctions were then cut into a bar shape with a width of 0.6 mm, thickness of 0.5 mm, and total length of 24.0 mm, where the length of each material is 12.0 mm. For the LIT measurements, the junctions were arranged in parallel with a gap of ~0.8 mm in the focal plane of the infrared camera, electrically connected in series, and suspended in air to prevent heat loss [Figure 1(a)]. To unify and enhance the infrared emission, the top surfaces of the junctions were coated with an insulating black ink with the infrared emissivity of >0.94. The calibration process to convert the observed infrared radiation intensity into temperature is detailed in [19]. The LIT measurements were performed while applying $\mathbf{J}_c$ with a square-wave amplitude of 1 A and frequency of $f$ = 0.5-12.5 Hz to the junctions at room temperature, where the direction of $\mathbf{J}_c$ was fixed in all the junctions [Figure 1(a)]. We also measured $D$, $\kappa$, $S$, and $\sigma$ of the materials by conventional methods (Table 1); the $\sigma$ values of all the materials and the parameters of Cu were used for the LIT-based thermal analyses, while the remaining parameters were used for the comparison to confirm the validity of the proposed method.

## 3. Results and discussion

Figure 2 shows the $A_\Pi$, $\phi_\Pi$, $A_J$, and $\phi_J$ images and corresponding line profiles averaged over the center area of the samples at $f$ = 10 Hz, where the target materials (Ni, Ni$_{95}$Pt$_5$, Fe, and Ti) and the reference material (Cu) are on the left and right halves, respectively. In the LIT measurement 1, clear current-induced temperature modulation signals appear in the vicinity of the junctions [$A_\Pi$ and $\phi_\Pi$ images in Figures 2(a) and 2(b)]. As shown in the line profiles, the $A_\Pi$ ($\phi_\Pi$) values exhibit asymmetrical exponential decay (linear change) with respect to the distance from the junctions. These behaviors are consistent with Equations (2) and (3), indicating that the signals in the $A_\Pi$ and $\phi_\Pi$ images originate from the oscillating $Q_\Pi$ at the junctions. The continuity of $A_\Pi$ and $\phi_\Pi$ at the junctions ensures that the interfacial thermal resistance is negligibly small. Importantly, the $\phi_\Pi$ values at the Ni/Cu and Ni$_{95}$Pt$_5$/Cu junctions differ by 180° from those at the Fe/Cu and Ti/Cu junctions, indicating that the sign of the temperature modulation for the former junctions is opposite to that for the latter junctions. This result is consistent with the fact that the relative Peltier coefficients for the Ni/Cu and Ni$_{95}$Pt$_5$/Cu junctions are opposite in sign to those for the Fe/Cu and Ti/Cu junctions (Table 1). The weak $A_\Pi$ decay and small $\phi_\Pi$ slope in Cu can be explained by the large $D$ of Cu. Qualitatively similar results were obtained at various values of $f$ (Figure S1 in the supplementary material). On the other hand, in the $A_J$ and $\phi_J$ images, almost uniform temperature modulation signals appear in the bulk of the materials; no signals at the junctions confirm that the Joule-heating contribution is extracted without the contamination by the Peltier contribution. As shown in Figure 2(c), the largest (smallest) Joule-heating signals were observed in Ti (Cu), consistent with Equation (4) and the parameters in Table 1. It is worth mentioning that the non-uniformity of $A_J$ and $\phi_J$ can be seen near the junctions due to the heat leakage between the materials. Such artifacts are suppressed with increasing $f$ (Figure S2 in the supplementary material). Thus, the accuracy of the LIT-based thermal analyses is improved at higher $f$. We also confirmed that, when $f > 5$ Hz, the heat loss at both the ends of the samples as well as the natural convention and thermal radiation are negligible in the LIT images (see the supplementary material).

Figures 3(a)-3(d) respectively show the $f$ dependence of $D$, $\kappa$, $S$ ($\Pi$), and $ZT$ at $T = 300$ K for the target materials, obtained by analyzing the LIT images. The obtained $D_A$ and $D_\phi$ values show notable deviation at low $f$ because of the heat loss [the inset to Figure 3(a)]. The heat loss effects can be minimized by taking the geometric mean $D = \sqrt{D_A D_\phi}$ [30], which is almost independent of $f$ and in good agreement with the values measured by the conventional method [Figure 3(a)]. The $\kappa$ values obtained by substituting $A_J^{\mathrm{TE}}$, $D$, and $\sigma$ into Equation (4) are shown in Figure 3(b); although the heat loss effects are unavoidable at low $f$, the results for $f > 5$ Hz are consistent with the values in Table 1 and the literature [19,31–33]. Figures 3(c) and 3(d) show $S$ (and the corresponding $\Pi$) and $ZT$ at $T = 300$ K for the materials, respectively. The consistency of the magnitude and sign of the coefficients confirms that the LIT-based method enables simultaneous estimation of thermal and thermoelectric properties. We also note that the $\rho c_p$ values estimated from the LIT-based analyses are also consistent with the values measured by the conventional methods (Figure S3 in the supplementary material).

To confirm the versatility of the LIT-based method, we performed the same measurements and thermal analyses with applying a magnetic field **H** (with the magnitude $H$) to the samples. Although the $H$ dependence of the electron transport coefficients is small in typical metals, $\sigma$, $\kappa$, and $S$ ($\Pi$) of ferromagnetic metals depend on the magnetization direction due to the anisotropic magnetoresistance, anisotropic magneto-thermal resistance, and anisotropic magneto-Seebeck (Peltier) effects, respectively [22]. Here, we show that the LIT-based method can be used for the simultaneous measurements of the anisotropic magneto-thermal resistance and anisotropic magneto-Seebeck and Peltier effects for many materials. We took the anisotropic magnetoresistance, i.e., the $H$ dependence of $\sigma$, into account in the following analyses but found that its contribution is negligibly small. In the experiments, the uniform **H** was applied along the longitudinal direction of the samples by using an electromagnet. The highest intensity of **H** is $\mu_0 H = 150$ mT with $\mu_0$ being the vacuum permeability, which is large enough to saturate the magnetization of all the bar-shaped ferromagnets [21].

Figure 4(a) shows the $H$ dependence of $S$, obtained from the LIT-based method. The small but finite $H$ dependence of $S$ was observed to appear in Ni and Ni$_{95}$Pt$_5$, while no clear dependence was observed in Fe and Ti. The $H$-dependent change ratio of $S$ for Ni$_{95}$Pt$_5$ is larger than that for Ni [Figure 4(d)], consistent with the previous report on the anisotropic magneto-Seebeck and Peltier effects [19,21]. We also found that the $H$ dependence of $\kappa$ for Ni and Ni$_{95}$Pt$_5$ is larger than that for Fe and Ti [Figures 4(b) and 4(e)]. The resultant $H$ dependence of $ZT$ calculated based on Equation (1) is shown in Figures 4(c) and 4(f). Ni$_{95}$Pt$_5$ exhibits the largest $ZT$ change, while Fe and Ti show almost no $H$ dependence; the observed $H$-dependent change ratio of $ZT$ at $\mu_0 H = 150$ mT are $-10.5\%$, $-16.8\%$, $1.5\%$, and $-0.4\%$ for Ni, Ni$_{95}$Pt$_5$, Fe, and Ti, respectively [Figure 4(f)]. In the materials used in this study, the magnetic field or magnetization dependent effects do not contribute to the enhancement of the thermoelectric performance. Nevertheless, the LIT-based method is useful because it enables simultaneous measurements of the $H$ dependence of $S$ and $\kappa$ for many samples.

Finally, we discuss the applicability limit of the LIT-based $ZT$ measurement method. The same LIT measurements and thermal analyses are applicable not only to metals but also to semiconductors when the interfaces comprising thermoelectric and reference materials exhibit Ohmic contact and negligible interfacial thermal resistance. In contrast, if a sample has large interfacial thermal resistance and exhibits discontinuous temperature modulation at the interface, the same analyses are not valid. The generalization of the LIT-based method to a system with finite interfacial thermal resistance is a remaining task for future studies.

## 4. Conclusion

We have proposed and demonstrated a *ZT* measurement method via the noncontact thermal imaging based on LIT. In this method, *D*, *κ*, *S* (Π), and *ZT* of multiple materials can be simultaneously estimated from the thermal analyses of the transient temperature distributions induced by the Peltier effect and Joule heating, enabling systematic, high-throughput, and versatile investigations of the thermal and thermoelectric properties. The parameters obtained by the LIT-based method are consistent with the values obtained by the conventional methods. This method can also be used for investigating the magnetic field or magnetization dependence of the thermal and thermoelectric transport properties. Although we measured the electrical conductivity separately from LIT, it is easy to integrate four-probe electrical measurements into a LIT system; by using such an integrated system, all the transport properties associated with *ZT* can be measured with a single apparatus. Thus, we anticipate that the LIT-based *ZT* measurement method will accelerate materials science research in thermoelectrics as well as spin caloritronics [22,34,35].

## Acknowledgments

The authors thank H. Ohta and T. Yamazaki for valuable discussions.

## Disclosure statement

The authors have no conflicts of interest directly relevant to the content of this article.

## Data availability statement

The data that support the findings of this study are available from the corresponding author upon reasonable request.

## Funding

This work was supported by CREST "Creation of Innovative Core Technologies for Nano-enabled Thermal Management" (JPMJCR17I1) from JST, Japan. A.M. was supported by JSPS through Research Fellowship for Young Scientists (18J02115).

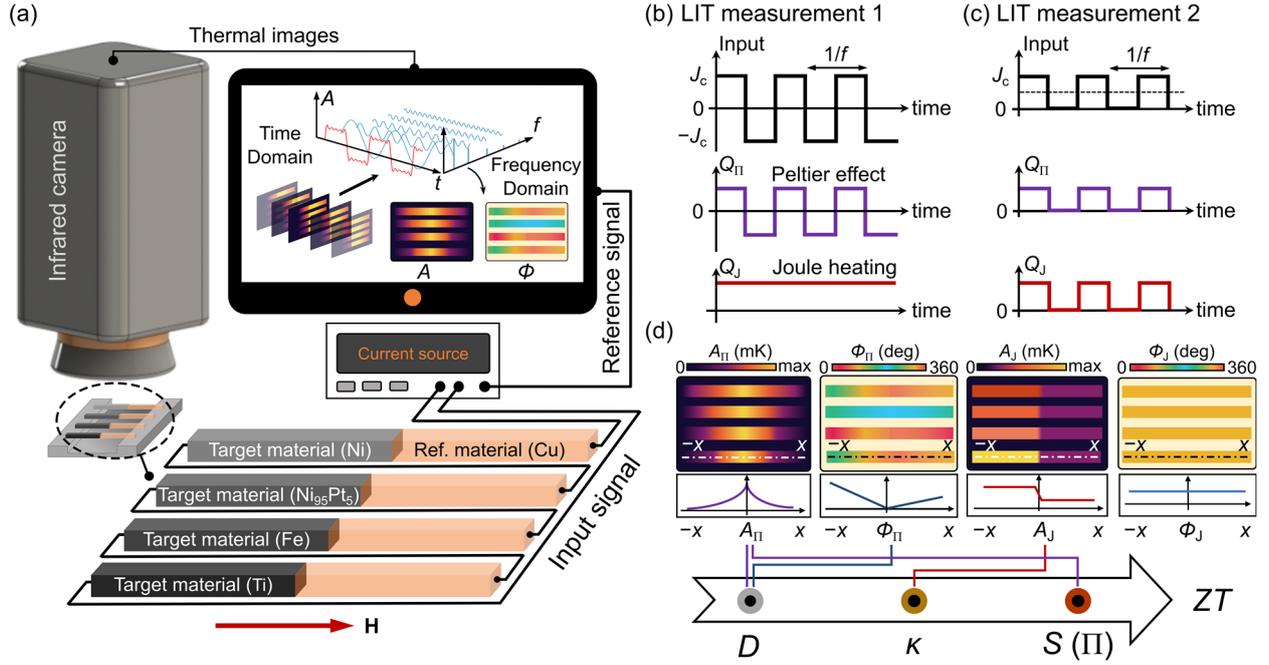

Figure 1. (a) Schematic of the LIT-based $ZT$ measurement technique. The samples are the bar-shaped junctions consisting of a thermoelectric material (target material; Ni, Ni$_{95}$Pt$_5$, Fe, and Ti) and a reference material (Cu). **H** denotes the external magnetic field. (b), (c) LIT conditions for measuring the temperature distribution induced by the Peltier effect (LIT measurement 1) and by the Joule heating (LIT measurement 2). In the LIT measurement 1 (2), a square-wave AC current with the frequency $f$, amplitude $J_c$ ($J_c/2$), and zero offset (finite DC offset $J_c/2$) is applied to the junctions. $Q_\Pi$ ($Q_J$) denotes heat release or absorption due to the Peltier effect (heat release due to the Joule heating). (d) Sequence of obtaining the thermal diffusivity $D$, thermal conductivity $\kappa$, Seebeck coefficient $S$ (Peltier coefficient $\Pi$), and $ZT$ by thermal analyses of LIT images. $A_\Pi$ and $\phi_\Pi$ ($A_J$ and $\phi_J$) denote the lock-in amplitude and phase of the temperature modulation induced by the Peltier effect (Joule heating), respectively.

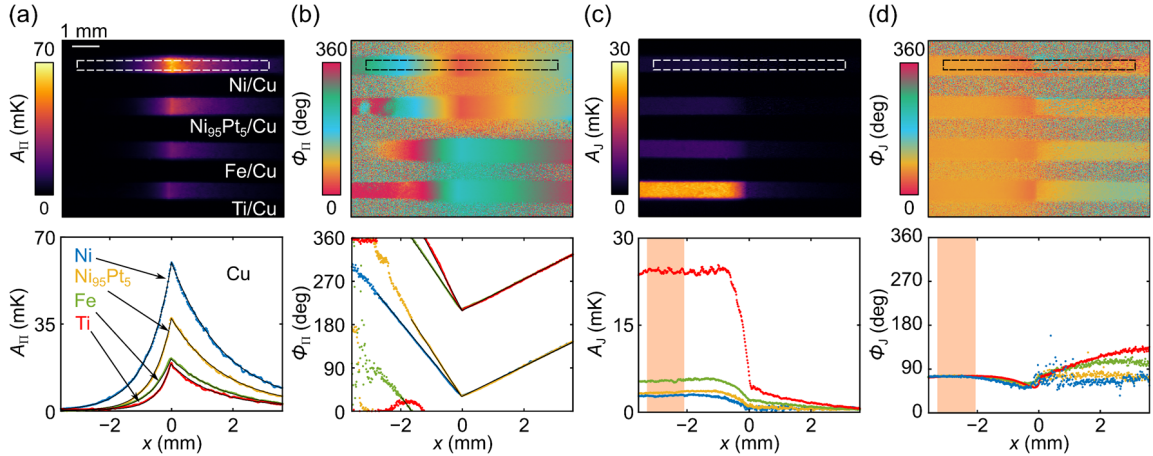

Figure 2. (a) $A_\Pi$, (b) $\phi_\Pi$, (c) $A_J$, and (d) $\phi_J$ images and corresponding line profiles along the longitudinal direction ($x$ direction) for the Ni/Cu, Ni$_{95}$Pt$_5$/Cu, Fe/Cu, and Ti/Cu samples at $f = 10$ Hz and $J_c = 1$ A. In each image, the junctions are placed around the center; the left (right) half in the images and line profiles corresponds to the target materials (reference material). The line profiles were obtained by taking the average of the raw profiles over the center areas (dashed rectangles) of the samples. $x = 0$ in the line profiles was determined by the peak position of $A_\Pi$ for each sample. The solid black lines in (a) and (b) represent the fitting results using Equations (2) and (3), respectively. The LIT data shown in this figure were measured in the absence of an external magnetic field.

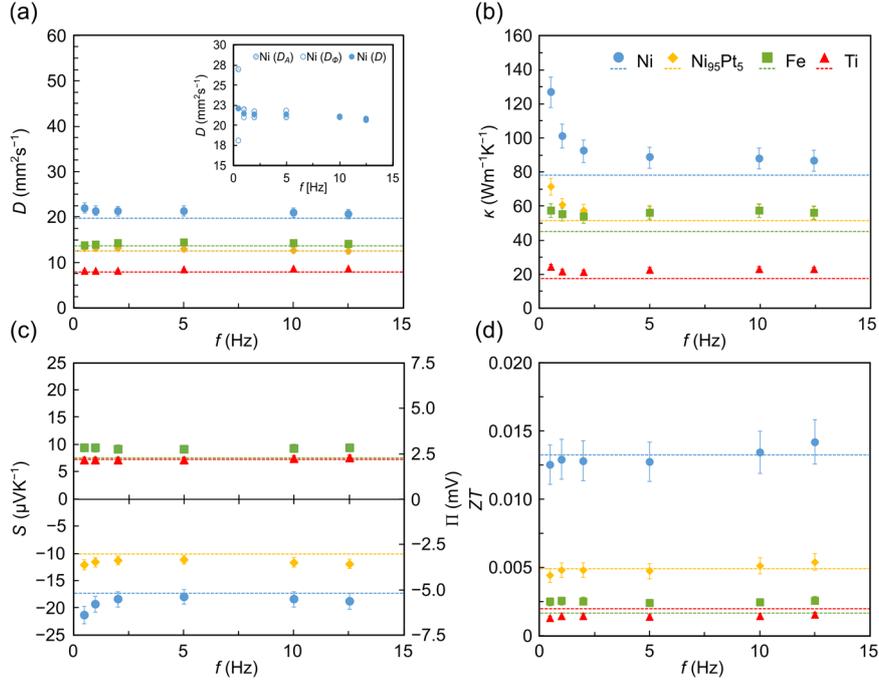

Figure 3. $f$ dependence of (a) $D$, (b) $\kappa$, (c) $S$ (and the corresponding $\Pi$), and (d) $ZT$ at $T = 300$ K for the Ni (blue circle), Ni$_{95}$Pt$_5$ (yellow diamond), Fe (green square), and Ti (red triangle) samples. $D$ was obtained from the geometric mean $D = \sqrt{D_A D_\phi}$ with the thermal diffusivity $D_A$ obtained from the $A_\Pi$ image and $D_\phi$ obtained from the $\phi_\Pi$ image by fitting. To estimate $\kappa$, we averaged the $A_J$ values over the orange rectangle areas in Figure 2, where the slope of the Joule-heating signal is negligibly small for $f > 5$ Hz (Figure S2 in the supplementary material). The dashed lines represent the values obtained by the conventional methods, where $D$ was measured by the laser flash method, $\kappa$ was estimated from $D$, specific heat capacity measured by the differential scanning calorimetry, and density measured by the Archimedes method, and $S$ and $\sigma$ were measured by using Seebeck Coefficient/Electric Resistance Measurement System (ZEM-3, ADVANCE RIKO, Inc.).

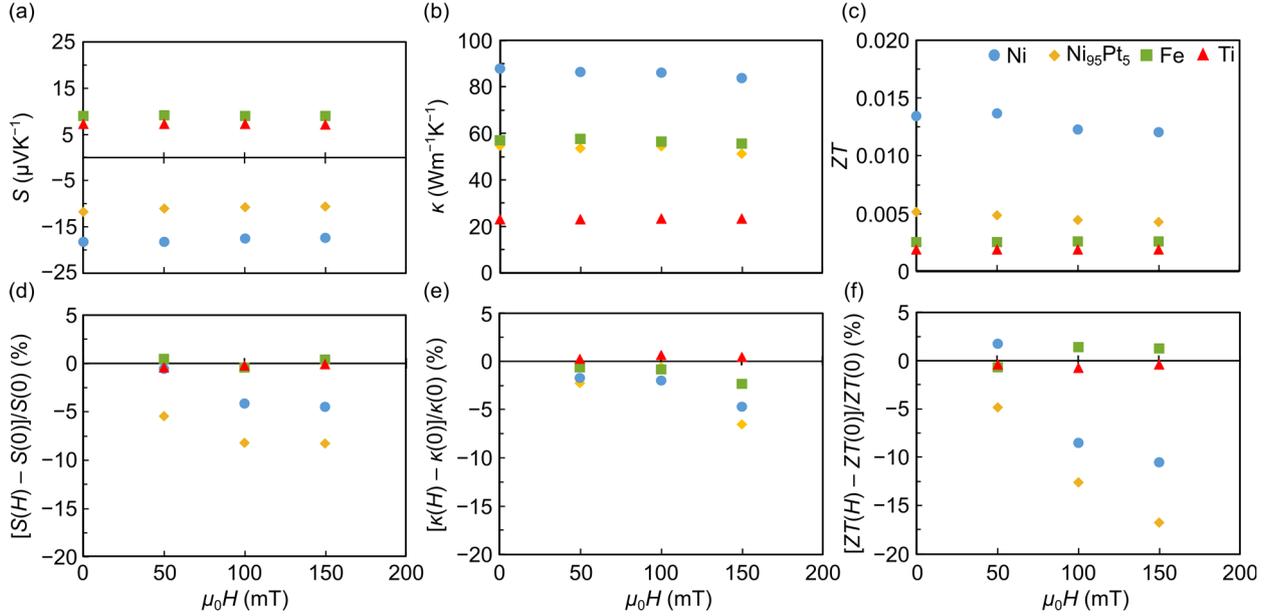

Figure 4. (a)-(c) $H$ dependence of $S$, $\kappa$, and $ZT$ at $T = 300$ K for the Ni (blue circle), Ni$_{95}$Pt$_5$ (yellow diamond), Fe (green square), and Ti (red triangle) samples. (d)-(f) $H$ dependence of $[S(H) - S(0)]/S(0)$, $[\kappa(H) - \kappa(0)]/\kappa(0)$, and $[ZT(H) - ZT(0)]/ZT(0)$. $S(0)$, $\kappa(0)$, and $ZT(0)$ denote the parameters at $\mu_0 H = 0$ mT with $\mu_0$ being the vacuum permeability. The data shown in this figure were estimated from the LIT images at $f = 10$ Hz.

Table 1. $D$, $\kappa$, $S$, and $\sigma$ of the materials measured by the conventional methods in the absence of a magnetic field.

| Material | $D$ (mm$^2$s$^{-1}$) | $\kappa$ (Wm$^{-1}$K$^{-1}$) | $S$ (µVK$^{-1}$) | $\sigma$ (Sm$^{-1}$) |
|---|---|---|---|---|
| Cu | 123.2 | 425.1 | 2.5 | $57.8 \times 10^6$ |
| Ni | 19.8 | 78.1 | −17.2 | $11.7 \times 10^6$ |
| Ni$_{95}$Pt$_5$ | 12.6 | 44.9 | −10.1 | $7.2 \times 10^6$ |
| Fe | 13.7 | 51.4 | 7.4 | $5.3 \times 10^6$ |
| Ti | 7.9 | 17.6 | 7.6 | $2.0 \times 10^6$ |

# Supplementary material

## S1. Lock-in frequency dependence of temperature modulation and heat loss

Figure S1 shows the $A_\Pi$ and $\phi_\Pi$ images and corresponding line profiles for various values of $f$ for the Ni/Cu, Ni$_{95}$Pt$_5$/Cu, Fe/Cu, and Ti/Cu samples at $\mu_0 H = 0$ mT. The clear temperature modulation induced by the Peltier effect appears at all the $f$ values but its spatial distribution depends on $f$. Because the temperature broadening due to thermal diffusion is suppressed by increasing $f$, the temperature modulation at higher $f$ values exhibits a sharper peak around the junctions. The $f$-dependent temperature broadening is characterized by the thermal diffusion length $\Lambda = [D/(\pi f)]^{0.5}$. Thus, the effect of the heat leakage at the ends of the samples can be reduced by increasing $f$, as discussed in the main text. In contrast, the magnitude of the temperature modulation monotonically decreases with increasing $f$; when the Peltier coefficient of a target material is small, one may have to decrease $f$ to improve the signal-to-noise ratio. The $f$ value should be determined by balancing these two factors.

The $f$ dependence of the temperature modulation induced by Joule heating is shown in Figure S2. The appropriate $f$ value for the LIT-based thermal analyses can be determined in a similar manner to the case of the temperature modulation induced by the Peltier effect.

Here, we discuss the effects of heat loss on the accuracy of the measurements. In general, heat transfer from the samples occurs via the thermal radiation $Q_{\text{rad}}$, convection $Q_{\text{conv}}$, and conduction $Q_{\text{cond}}$ [1,2]. Because of the small temperature modulation observed in this study, the effect of $Q_{\text{rad}}$ and $Q_{\text{conv}}$ can be estimated via $Q_{\text{rad}} + Q_{\text{conv}} \propto h(T - T_{\text{amb}})$, where $h = h_{\text{rad}} + h_{\text{conv}}$ is the combined heat transfer coefficient, $T$ the sample temperature, and $T_{\text{amb}}$ the ambient temperature [1,3]. At these conditions, $h_{\text{rad}} = 4\varepsilon\delta T_{\text{amb}}^3$ gives a maximum value of ~6 Wm$^{-2}$K$^{-1}$ for the extreme case of a blackbody (emissivity $\varepsilon = 1$), where $\delta$ is the Stefan-Boltzmann constant. On the other hand, $h_{\text{conv}}$ at lowest $f$ shown in Figure S2 is estimated to be ~3 Wm$^{-2}$K$^{-1}$ for a horizontal slab [4], resulting in the total $h$ of ~9 Wm$^{-2}$K$^{-1}$. In the measurements of the Peltier effect, the heat loss ratio can be evaluated by $(Q_{\text{rad}} + Q_{\text{conv}})/Q_\Pi \sim (h/\kappa d)\Lambda^2$, where $\kappa$ and $d$ are the thermal conductivity and thickness of the slab, respectively [3]. On the other hand, the heat loss ratio in the measurements of Joule heating can be evaluated by $(Q_{\text{rad}} + Q_{\text{conv}})/Q_J \sim h/(\rho c_p d)$. Based on these assumptions and the parameters measured by the conventional method (Table 1), we confirmed that the heat loss ratios are always less than 1% in the measurement $f$ range. The heat loss due to $Q_{\text{cond}}$ in our experimental setup represents the heat leaked from both the ends of the samples to the reference material and the electrical connection since the samples are suspended in air. In the measurements of the Peltier effect, the $Q_{\text{cond}}$ contribution can be neglected if $\Lambda$ is shorter than the distance between the junction and the region of interest, which can be controlled by changing $f$ [3]. In the same manner, the effect of $Q_{\text{cond}}$ on the Joule heating measurements can be neglected at high $f$. By using the $D$ values in Table 1, we estimated $\Lambda$ at $f = 5$ Hz to be 1.1, 0.9, 0.9, and 0.7 mm for Ni, Ni$_{95}$Pt$_5$, Fe, and Ti, respectively, much less than the length of the sample. This estimation confirms that the LIT-based thermal analyses are not affected by $Q_{\text{cond}}$ for $f > 5$ Hz. Thus, we selected the data at $f = 10$ Hz for Figures 2 and 4 in the main text.

## S2. Volumetric specific heat capacity measurement

An additional merit of our LIT-based method is that the volumetric specific heat capacity $\rho c_p$ of the target materials can be estimated at the same time through the relation $\rho c_p = \kappa/D$. Figure S3(a) shows the $f$ dependence of $\rho c_p$ of the target materials. In a similar manner to the other parameters, the estimated $\rho c_p$ values for $f > 5$ Hz are consistent with the values obtained by the conventional measurement methods. This result indicates that the LIT-based method is also capable of simultaneous non-contact measurements of $\rho c_p$ for many materials. Furthermore, the LIT-based method allows us to measure the $H$ dependence of $\rho c_p$ [Figure S3(b)]. Unlike the other transport coefficients, the $\rho c_p$ values for the materials used in this study show no clear $H$ dependence.

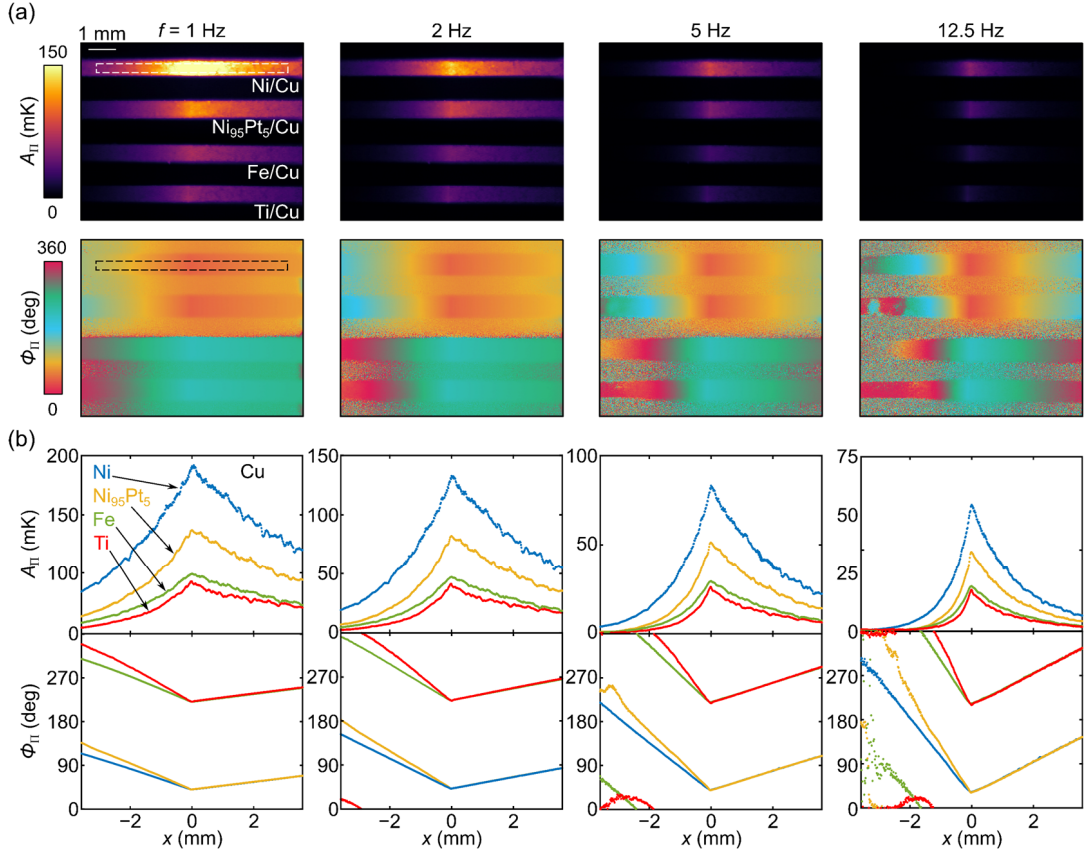

Figure S1. (a) $A_\Pi$ and $\phi_\Pi$ images and (b) corresponding line profiles for the Ni/Cu, Ni$_{95}$Pt$_5$/Cu, Fe/Cu, and Ti/Cu samples for various values of $f$ at $J_c$ = 1 A and $\mu_0 H$ = 0 mT. The signals shown in this figure are attributed to the Peltier effect at the junctions. The line profiles were obtained by taking the average of the raw profiles over the center areas (dashed rectangles) of the samples.

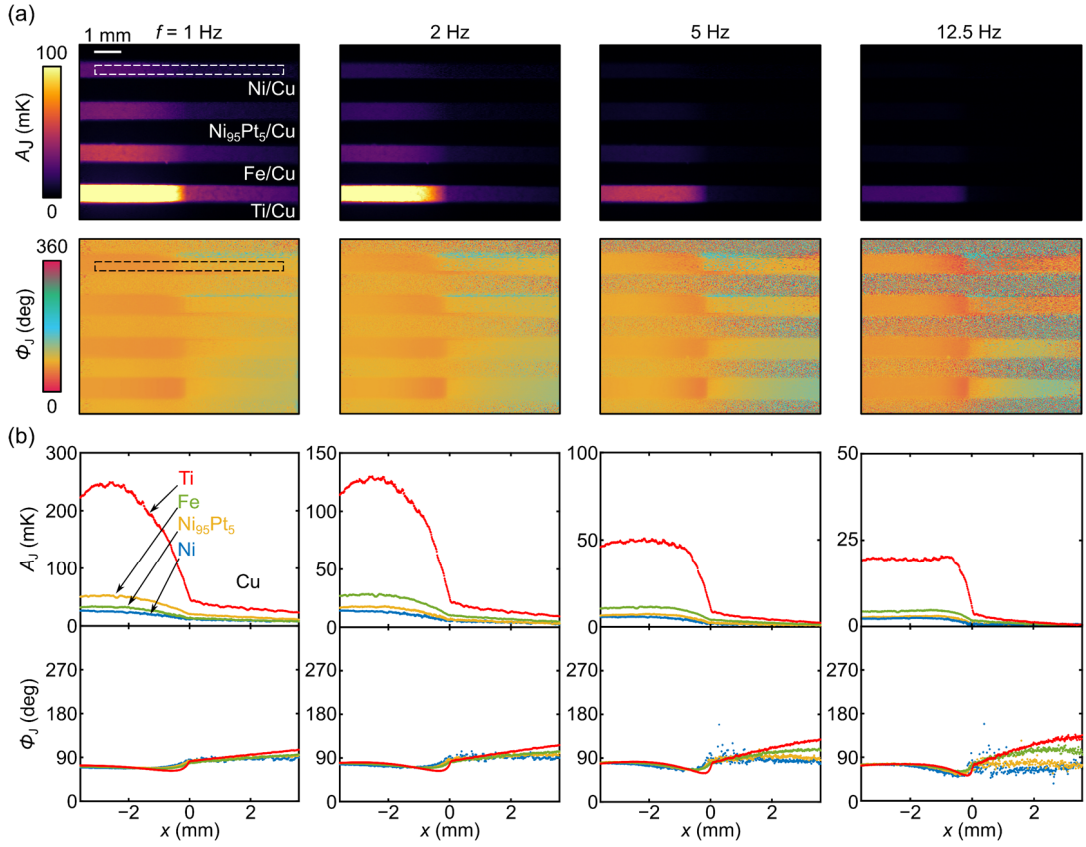

Figure S2. (a) $A_J$ and $\phi_J$ images and (b) corresponding line profiles for the Ni/Cu, Ni$_{95}$Pt$_5$/Cu, Fe/Cu, and Ti/Cu samples for various values of $f$ at $J_c$ = 1 A and $\mu_0 H$ = 0 mT. The signals shown in this figure are attributed to the Joule heating in the bulk of the materials.

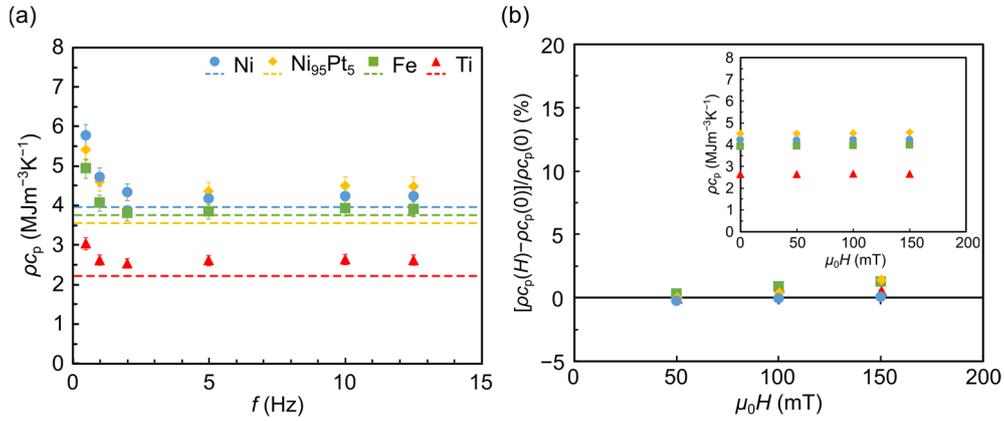

Figure S3. (a) $f$ dependence of $\rho c_p$ for the Ni (blue circle), Ni$_{95}$Pt$_5$ (yellow diamond), Fe (green square), and Ti (red triangle) samples, estimated from the LIT images at $\mu_0 H$ = 0 mT. The dashed lines represent the values estimated from the specific heat capacity measured by the differential scanning calorimetry and density measured by the Archimedes method. (b) $H$ dependence of $[\rho c_p(H) - \rho c_p(0)]/\rho c_p(0)$, estimated from the LIT images at $f$ = 10 Hz. $\rho c_p(0)$ refers to the volumetric specific heat capacity at $\mu_0 H$ = 0 mT. The inset to (b) shows the $H$ dependence of $\rho c_p$.